\documentclass[runningheads]{llncs}
\usepackage[T1]{fontenc}

\usepackage{mathtools}
\usepackage{graphicx}

\usepackage{amsmath}
\usepackage{bm}

\usepackage[T1]{fontenc}
\usepackage{multirow}
\usepackage{graphicx}
\usepackage{amssymb}
\usepackage{amsfonts}
\usepackage{booktabs}
\usepackage{array}

\newcolumntype{L}[1]{>{\raggedright\let\newline\\\arraybackslash\hspace{0pt}}m{#1}}
\newcolumntype{C}[1]{>{\centering\let\newline\\\arraybackslash\hspace{0pt}}m{#1}}
\newcolumntype{R}[1]{>{\raggedleft\let\newline\\\arraybackslash\hspace{0pt}}m{#1}}

\begin{document}
\title{Atlas-powered deep learning (ADL) - \\ application to diffusion weighted MRI}
\titlerunning{Atlas-powered deep learning}
%
\author{ Davood Karimi and Ali Gholipour }
%
%
\institute{ Computational Radiology Laboratory of the Department of Radiology at Boston Children's Hospital, and Harvard Medical School, Boston, Massachusetts, USA
}
\maketitle              
\begin{abstract}

Deep learning has a great potential for estimating biomarkers in diffusion weighted magnetic resonance imaging (dMRI). Atlases, on the other hand, are a unique tool for modeling the spatio-temporal variability of biomarkers. In this paper, we propose the first framework to exploit both deep learning and atlases for biomarker estimation in dMRI. Our framework relies on non-linear diffusion tensor registration to compute biomarker atlases and to estimate atlas reliability maps. We also use nonlinear tensor registration to align the atlas to a subject and to estimate the error of this alignment. We use the biomarker atlas, atlas reliability map, and alignment error map, in addition to the dMRI signal, as inputs to a deep learning model for biomarker estimation. We use our framework to estimate fractional anisotropy and neurite orientation dispersion from down-sampled dMRI data on a test cohort of 70 newborn subjects. Results show that our method significantly outperforms standard estimation methods as well as recent deep learning techniques. Our method is also more robust to stronger measurement down-sampling factors. Our study shows that the advantages of deep learning and atlases can be synergistically combined to achieve unprecedented accuracy in biomarker estimation from dMRI data.


\end{abstract}

\section{Introduction}

Diffusion weighted magnetic resonance imaging (dMRI) is the de-facto imaging modality for probing the brain micro-structure in vivo. Biomarkers estimated with dMRI are widely used to study normal and abnormal brain development \cite{alexander2019imaging,huppi2006diffusion}. Estimation of these biomarkers from the dMRI signal entails solving inverse problems that can range from non-linear least squares to complex non-convex optimization problems \cite{novikov2019quantifying,harms2017robust}. Accurate estimation depends on dense high-quality measurements, which may be difficult or impossible to obtain in many applications such as neonatal and pediatric studies.

Machine learning, and in particular deep learning (DL), has emerged as a potent alternative to classical optimization-based estimation methods for dMRI analysis. Rather than solving the inverse problem directly, DL methods aim to learn the complex relation between the dMRI signal and the parameter of interest from a set of training data. Recent studies have used DL models for estimating diffusion tensor, diffusion kurtosis, multi-compartment models, and fiber orientation distribution, to name just a few \cite{golkov2016,gibbons2019simultaneous,karimi2021learning,tian2020deepdti,li2021superdti}. They have shown that DL methods can accurately estimate dMRI biomarkers using far fewer measurements than optimization-based techniques.

Statistical atlases are models of expected anatomy and anatomical variation. They are routinely used in medical image analysis to characterize normal anatomy and for identifying anatomical differences between normal and/or abnormal subjects or populations. In neuroimaging, study-specific atlases offer higher sensitivity and specificity of analysis than off-the-shelf atlases. Several prior works have used dMRI atlases to study brain development and maturation \cite{khan2019fetal,akazawa2016probabilistic}, to assess the impact of neurological diseases \cite{oishi2009atlas,hasan2014serial}, and for various other purposes \cite{hagler2009automated,saghafi2017spatio}.

However, no prior work has attempted to leverage DL and atlases for dMRI biomarker estimation in a unified framework. Such a framework makes much intuitive sense. While advanced DL methods are very effective in learning the complex mapping between the dMRI signal and the biomarker of interest, atlases can supply additional useful information that may be absent from the local diffusion signal. The information contained in the atlas can be particularly useful where the local diffusion signal is not adequate for accurate estimation, such as when the number of measurements is low or the signal is noisy. Some studies have shown that atlases can improve the performance of DL methods for segmentation (e.g., \cite{oguz2018combining,diniz2020esophagus}). There have also been efforts to use atlases or other sources of prior information such as distribution of noise and parameter values with classical estimation methods \cite{veraart2013comprehensive}. For example, Taquet et al. showed that estimation accuracy for diffusion compartment models can be improved by using a population-informed prior in a graphical model \cite{taquet2015improved}. Anderson \cite{andersson2008maximum} used probabilistic priors on model parameters in a maximum likelihood estimation framework for diffusion tensor estimation. Also using a Bayesian estimation approach, Clayden et al. \cite{clayden2016microstructural} found that prior information was very influential for estimating axon radii from dMRI signal. In another Bayesian work for NODDI estimation, Mozumder et al. \cite{mozumder2019population} incorporated the priors learned on 35 subjects. However, to the best of our knowledge, no prior work has used atlases within a DL framework for dMRI biomarker estimation.

On the other hand, designing such a framework is not trivial. An atlas can only represent the average of a population; it lacks the subtle but important variations among the individuals in the population. Furthermore, the correspondence between an atlas and an individual subject's brain is complex and spatially-varying. The atlas may match one subject much better than another subject. Within the brain of a subject, the match between the atlas and the subject depends on the local anatomy. It is not clear how these information can be incorporated into a machine learning framework.

In this work, we propose a framework that brings together the power of DL models and atlases for dMRI biomarker estimation. Our framework nicely addresses the challenges outlined above by providing methods to compute the reliability of the atlas and its degree of correspondence with a subject in a spatially-varying manner. We use our proposed framework to estimate fractional anisotropy and neurite orientation dispersion from down-sampled dMRI data of newborn subjects and show that it is significantly more accurate than standard estimation methods as well as recent DL techniques.

\section{Materials and methods}

\subsection{Data}

We used 300 dMRI scans from the Developing Human Connectome Project (DHCP) dataset \cite{bastiani2019}. Each scan is from a different subject. We used 230 of the scans for model development, including atlas creation and DL model training. We used the same 230 scans to train the competing DL models (described below). We used the remaining (completely independent) 70 scans to test our method and the competing methods. The age of the subjects at the time of scan ranged between 31 and 45 gestational-equivalent weeks. We have focused on this dataset because dMRI analysis for this age range is especially challenging due to higher free water content, incomplete myelination, and overall lower data quality \cite{dubois2014early}.

\subsection{Atlas development}

Because of rapid brain development in the neonatal period, a single atlas cannot represent the entire age range \cite{pietsch2019framework,uus2021multi}. Therefore, we built atlases of the biomarkers of interest separately at one-week intervals. To build atlases for GA of week 35, for example, we used subjects with GA between 34.5 and 35.5. For each GA, we used 10 scans to build the atlas. Our experience and prior works \cite{pietsch2019framework,khan2019fetal,uus2021multi} have shown that approximately 10 subjects are sufficient for each GA.

In this study, we focus on two biomarkers: 1) Fractional anisotropy (FA) from the diffusion tensor model, which is arguably the most widely used dMRI biomarker, and 2) Orientation dispersion (OD) from the NODDI model \cite{zhang2012noddi}, which is a more complex model than diffusion tensor and which has been shown to be a more specific biomarker in some studies \cite{andica2021neurite,palacios2020evolution}. 

\begin{figure}
\includegraphics[width=\textwidth]{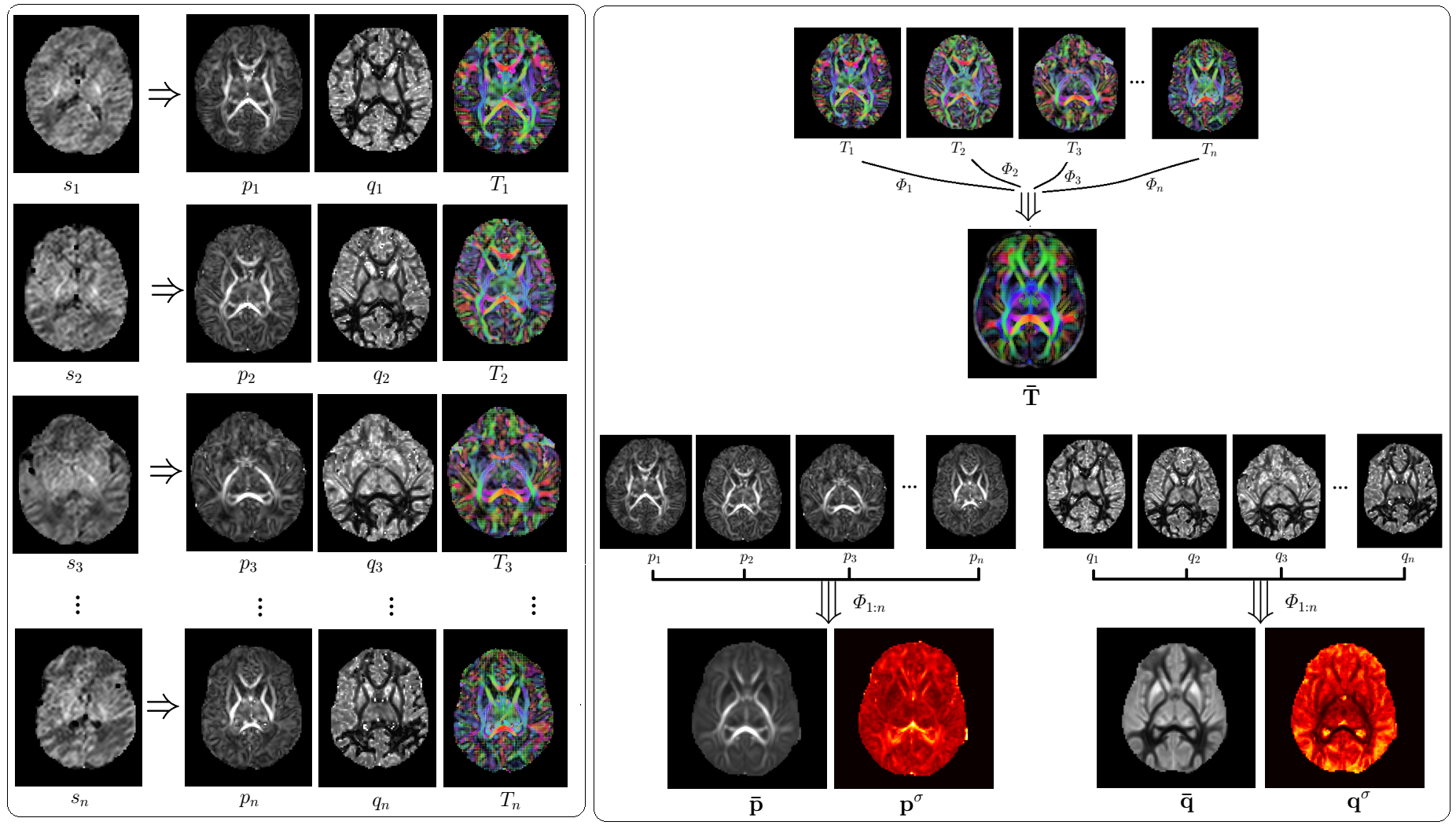}
\caption{Left: Biomarkers of interest and diffusion tensor are computed for each subject. Right: tensor-to-tensor registration is used to compute the atlas and atlas confidence maps for each biomarker.} \label{fig:atlas_creation}
\end{figure}

Given the dMRI signal volumes $\{ s_i \}_{i=1}^n$ for $n$ subjects, we compute the biomarkers of interest separately for each subject. Regardless of which biomarkers are explored, we also always compute the diffusion tensor for each subject because we use the tensors for accurate spatial alignment. This is shown in Figure \ref{fig:atlas_creation}, where we have used $p$ and $q$ to denote the biomarkers considered in this work and $T$ to denote the diffusion tensor. Given the set $\{ T_i \}_{i=1}^n$ of subject tensors, we compute a set of transformations  $\{ \Phi_i \}_{i=1}^n$ that align these tensors into a common atlas space and compute a mean tensor $\bar{T}$. This is done using an iterative approach that computes a series of rigid, affine, and diffeomorphic non-rigid registrations and updates $\bar{T}$ at every iteration, rather similar to common practice \cite{pietsch2019framework,khan2019fetal}. Specifically, we perform five iterations of rigid registration, followed by five iterations of affine registration, and finally ten iterations of non-rigid registration. The final registration transform $\Phi_i$ for subject tensor $T_i$ is the composition of the final affine and non-rigid transforms.

Using the transformations $\{ \Phi_i \}_{i=1}^n$ computed via tensor-to-tensor registration described above, we transform the biomarker maps into the common atlas space. We compute the mean of the transformed biomarker maps as the biomarker atlas. Furthermore, we compute the standard deviation of the transformed biomarker maps as a measure of atlas confidence. Formally, for biomarker $p$:

\begin{equation}
\bar{p}= \text{mean} \Big[ \Phi_i(p_i) \Big]_{i=1}^n , \hspace{5mm}  p^{\sigma}= \text{std} \Big[ \Phi_i(p_i) \Big]_{i=1}^n
\end{equation}

\noindent where mean and standard deviation are computed in a voxel-wise manner across subjects. Larger values of $p^{\sigma}$ indicate higher variability/disagreement between the biomarker maps of subjects used to create the atlas and, hence, \emph{lower} confidence. Figure \ref{fig:atlas_creation} shows example atlas and atlas confidence maps for FA and OD.

\subsection{Biomarker estimation for an individual subject}

Given the dMRI signal volume for an individual subject, $s_k$, we compute the desired biomarker(s) for that subject via the following steps, which have been shown for an example subject in Figure \ref{fig:subject_analysis}.

\subsubsection{Step 1: Atlas-to-subject alignement.}

In order to exploit the information encoded in an atlas, we need to accurately register it to the subject space. As in our atlas development described above, we use tensor-to-tensor registration for this alignement. Hence, regardless of the biomarker(s) being estimated, we compute the diffusion tensor, $T_k$, for the subject. We then compute affine+non-linear registration transforms that map the atlas tensor $\bar{T}$ to the subject tensor, $T_k$. We use $\Phi_k$ to denote the composition of these affine and non-linear transforms. Hence, $\Phi_k$ describes the complete spatial alignment from the atlas space to the subject. We denote the template tensor transformed to the subject space with $\bar{T}_k= \Phi_k(\bar{T})$.

The registration between the template tensor and the subject tensor is never perfect. The accuracy of this registration is a potentially important piece of information because it indicates where the prior information is more reliable. In other words, if at voxel $i$ the registration between $\bar{T}$ and $T_k$ is more accurate, then we have a higher incentive to trust the biomarker atlas at the location of that voxel. The accuracy of this registration is spatially varying and depends on at least three factors: 1) Accuracy of computation of $T_k$, which in turn depends on the quality of the diffusion signal, $s_k$, 2) Degree of similarity between the subject and the atlas, and 3) Accuracy of the registration procedure that aligns the atlas to the subject. We propose the following practical formulation to estimate the error of this alignment:

\begin{equation}   \label{eq:alignment_error}
\Phi_k^{\text{err}}= \theta \big( \bar{T}_k, T_k \big) . \exp \Big( - \min \big[ \text{FA}(\bar{T}_k) , \text{FA}(T_k) \big] / \tau \Big).
\end{equation}

This formulation has two terms. The first term, $\theta$, measures the angle between the major eigenvectors of $\bar{T}_k$ and $T_k$. Clearly, smaller angles indicate more accurate registration. The second term is introduced to down-weight the registration accuracy for the location of less anisotropic tensors such as gray matter and cerebrospinal fluid (CSF). For CSF, for example, the tensor is spherical and the computed orientation of the major eigenvector is not reliable but the eigenvectors of $T_k$ and $\bar{T}_k$ may be very close to each other by chance, hence artificially making $\theta$ very small. By using the minimum of the FAs, if either $\bar{T}_k$ or $T_k$ has a low anisotropy, the second term will have a larger value. We set $\tau=0.2$, which we found empirically to work well.

\subsubsection{Step 2: Estimation using a DL model.}

\begin{figure}
\includegraphics[width=0.9\textwidth]{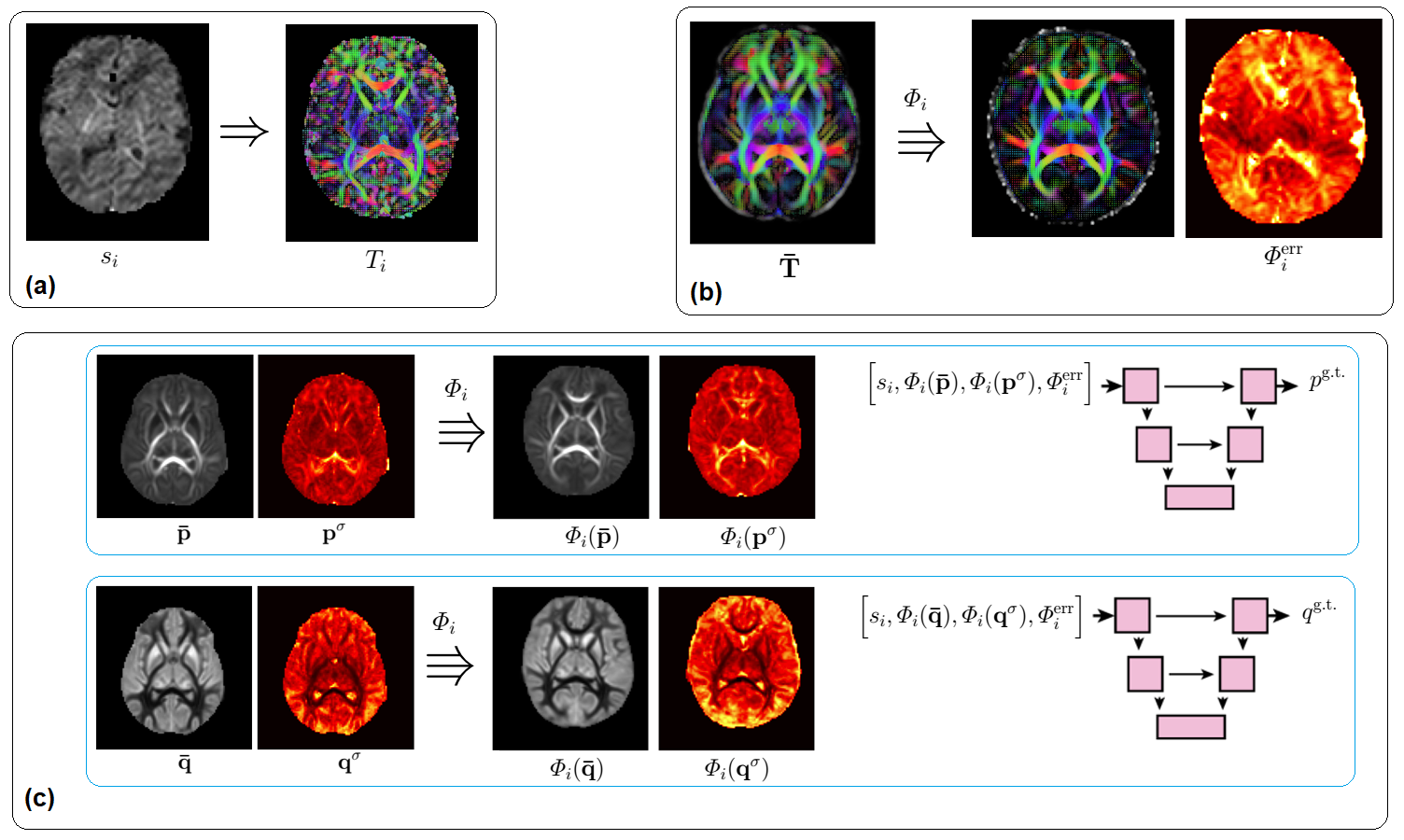}
\caption{Our proposed method for estimating dMRI biomarkers using atlases and DL. (a) We compute the diffusion tensor $T_k$  from the subject data. (b) We compute the registration $\phi_k$ from the atlas to the subject using tensor-to-tensor registration, and then the error of this registration $\Phi_k^{\text{err}}$ using Eq. \ref{eq:alignment_error}. (c) We align the biomarker atlases and atlas confidence maps to the subject using $\Phi_k$. We then feed these information, along with the dMRI signal $s_k$ to the DL model, which aims to predict the ground truth biomarker values.} \label{fig:subject_analysis}
\end{figure}

To compute a specific biomarker for an individual subject, standard techniques only use the diffusion signal, $s_k$. In our framework, we also utilize the prior information encoded in the biomarker atlas as described above. Specifically, for estimating a biomarker $p$ for subject $k$, we have three additional pieces of information; 1) the biomarker atlas registered to the subject space $\Phi_k(\bar{p})$, 2) the biomarker atlas confidence registered to the subject space $\Phi_k(\bar{p})$, and 3) registration error $\Phi_k^{\text{err}}$. Since these three pieces of information are spatially varying and aligned with the subject space, we simply concatenate them with the diffusion signal to generate the input to our DL model, as shown in Figure \ref{fig:subject_analysis}. For example, for estimating the biomarker $p$ for subject $k$, the input to the network is $\Big[ s_k, \Phi_k(\mathbf{ \bar{p} }), \Phi_k(\mathbf{ p^{\sigma}}) , \Phi_k^{\text{err}} \Big]$. The estimation target is the ground-truth parameter map, $p^{\text{g.t.}}$, computed as described below.

We used the UNet++ \cite{zhou2018unet} as our DL architecture. As it has been recently demonstrated for image segmentation in \cite{isensee2021nnu}, our experience shows that the exact network architecture is not critical. We used patches of size $48^3$ voxels. The number of network input channels was equal to the number of diffusion measurements plus 3, as described above. We train a separate network for each biomarker, hence the network had only one output channel to compute the scalar biomarker of interest. We set the number of feature maps in the first stage of the network to be 12, which was the largest possible on our GPU memory. For training, we sampled blocks from random locations in the training images. At test time, we used a sliding window with a stride of 16 voxels in each dimension to estimate the biomarker for an input dMRI volume of arbitrary size.

\subsubsection{Compared methods and evaluation strategy.}

For FA estimation, we compared our method with: 1) Constrained weighted linear least-squares (\textbf{CWLLS}) \cite{koay2006unifying}, which is the standard method; 2) \textbf{Deep-DTI} \cite{tian2020deepdti}, which is a recent DL method based on CNNs. This method exploits the anatomical T2 image, in addition to the diffusion signal, for estimation. Hence, for Deep-DTI we also used the T2 image, which we registered to the dMRI volume, and 3) \textbf{Super-DTI} \cite{li2021superdti}, which is a recent DL method based on CNNs. For OD estimation, we compared our method with: 1) \textbf{Dmipy} \cite{fick2019dmipy}, which follows a standard optimization-based estimation approach, 2) Microstructure Estimation using a Deep Network (\textbf{MEDN+}) \cite{ye2017tissue}, which is a DL method that has been inspired by AMICO \cite{daducci2015accelerated} and significantly outperforms AMICO too, and 3) Another recent CNN-based model \cite{gibbons2019simultaneous}, which we refer to as \textbf{CNN-NODDI}.

Each DHCP scan includes 20 $b=0$ measurements and 280 diffusion weighted measurements at $b=400$ (n=64), $b=1000$ (n=88), and $b=2600$ (n=128). For FA it is known that b values close to 1000 are optimal \cite{jones1999optimal}. Hence, we used all 88 measurements in the $b=1000$ shell (along with all b=0 measurements) to compute the ground truth using CWLLS. We then selected subsets of 12 and 6 measurements from this shell for each subject, which represent measurement down-sampling factors of approximately 7 and 15, respectively. To select the 6 measurements, similar to \cite{tian2020deepdti,karimi2022diffusion}, we considered the 6 optimal diffusion gradient directions proposed in \cite{skare2000condition} and chose the measurements that were closest to those directions. To select the 12 measurements, we selected these measurements to be close to uniformly spread on the sphere, as suggested in \cite{jones1999optimal,karimi2021learning}.

For OD (and NODDI), we used all 300 measurements to reconstruct the ground truth using Dmipy \cite{fick2019dmipy}. For DL-based reconstruction, prior works have typically used 20-60 measurements from more than one shell \cite{ye2017tissue,gibbons2019simultaneous}. Here, we choose either 6 and 15 measurements from each of the $b=1000$ and $b=2600$ shells, for a total of 12 and 30 measurements, which represent downsampling factors of approximately 24 and 10, respectively. We selected these measurements to be close to uniformly spread on the sphere, using an approach similar to \cite{karimi2021learning}. For a fair comparison, for both FA and OD we used the same down-sampled datasets for our method and for all competing techniques.

\subsubsection{Implementation and training.}

We used the DTI-TK software \cite{zhang2006deformable} to compute all registrations, including the registrations needed in building the atlases. In creating the biomarker atlases we used MRtrix \cite{tournier2019mrtrix3} to compute the diffusion tensors and FA, and we used Dmipy \cite{fick2019dmipy} to compute OD. We implemented all DL techniques in TensorFlow 1.14 under Python 3.7. We ran all algorithms on a Linux machine with 16 CPU cores and an NVIDIA GeForce GTX 1080 GPU.

We trained our model by minimizing the $\ell_2$ norm between the predicted and ground truth biomarker using Adam \cite{kingma2014}, a batch size of 10, and an initial learning rate of $10^{-4}$ that was reduced by half every time the validation loss did not decrease after a training epoch. For the competing methods, we followed the training procedures recommended in the original papers.

\section{Results and Discussion}

\subsection{Comparison with other techniques}

\begin{figure}
\includegraphics[width=1.0\textwidth]{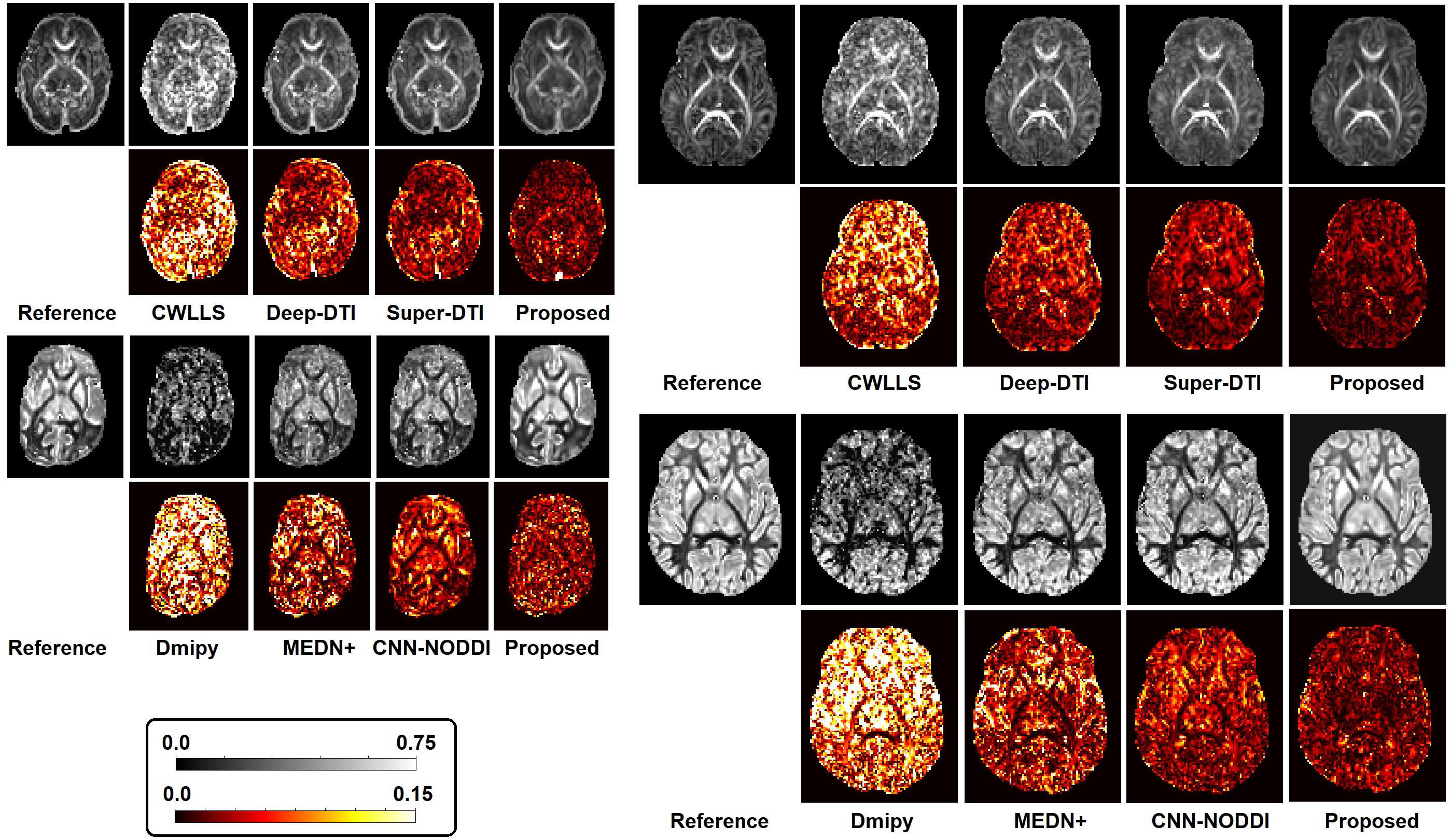} 
\caption{Example FA (top) and OD (bottom) reconstructions by different methods. In each of the four examples, the bottom row shows maps of absolute estimation error.} 
\label{fig:sample_results}
\end{figure}

Tables \ref{table:FA_results} and \ref{table:OD_results} show the reconstruction error for different methods for FA and OD, respectively, computed on the 70 independent test subjects. For both FA and OD, the DL methods were substantially more accurate than the standard optimization-based techniques (i.e., CWLLS and Dmipy). The proposed method achieved lower errors than the other DL methods for both FA and OD at both down-sampling factors. We used paired t-tests to compare our method with the other techniques. For both FA and OD and at both down-sampling factors, the estimation error for our method was significantly lower than the error for any of the compared techniques ($p<0.001$). Figure \ref{fig:sample_results} shows example reconstruction results for different techniques. Our method achieves lower errors across the brain for both FA and OD.

\begin{table*}[!htb]
\footnotesize
 \caption{\footnotesize{FA estimation errors for the proposed method and compared techniques.}}
  \label{table:FA_results}
   \begin{center}
    \begin{tabular}{ L{29mm}  C{19mm} C{21mm} C{21mm} C{22mm} }
\hline
no. of measurements & CWLLS & Deep-DTI & Super-DTI & Proposed   \\ \hline
$n=6$ & $0.111 \pm 0.014$ & $0.048 \pm 0.008$  & $0.048 \pm 0.005$ & $\bm{0.040 \pm 0.005}$   \\ 
$n=12$ & $0.053 \pm 0.007$ & $0.044 \pm 0.006$  & $0.043 \pm 0.004$ & $\bm{0.039 \pm 0.005}$   \\ 
\hline
\end{tabular}
  \end{center}
\end{table*}

\begin{table*}[!htb]
\footnotesize
 \caption{\footnotesize{Comparison of OD estimation errors for the proposed method and competing techniques.}}
  \label{table:OD_results}
   \begin{center}
    \begin{tabular}{ L{29mm}  C{19mm} C{21mm} C{21mm} C{22mm} }
\hline
no. of measurements & Dmipy & MEDN+ & CNN-NODDI & Proposed   \\ \hline
$n=16$ & $0.138 \pm 0.032$ & $0.064 \pm 0.028$  & $0.058 \pm 0.011$ & $\bm{0.047 \pm 0.004}$   \\ 
$n=30$ & $0.096 \pm 0.029$ & $0.052 \pm 0.030$  & $0.050 \pm 0.008$ & $\bm{0.044 \pm 0.005}$   \\ 
\hline
\end{tabular}
  \end{center}
\end{table*}

For both FA and OD, our proposed method showed a smaller increase in error as the down-sampling rate increased. Specifically, the FA estimation error for our method increased by 2.5\% as the number of measurements was reduced from 12 to 6, compared with 9\% and 12\% for Deep-DTI and Super-DTI, respectively. For OD, the estimation error for our method increased by 7\% as the number of measurements was decreased from 30 to 16, compared with 23\% and 16\% for MEDN+ and CNN-NODDI, respectively.

The training time for our model was 10 hours, compared with 10-60 hours for the other DL methods. To estimate FA or OD for a dMRI test volume, our model required $73 \pm 20$ seconds. Approximately 90\% of this time was spent on computing the atlas-to-subject registration. The average computation time for the other DL methods ranged from 10 seconds for SuperDTI to 30 minutes for MEDN+. The average computation times for CWLLS and Dmipy were, respectively, 20 seconds and 3.2 hours. Nonetheless, computation time for dMRI analysis is not a critical factor since fast estimation is typically not a requirement.


\subsection{Ablation experiments}

Table \ref{table:ablation} shows the results of some ablation experiments. We have performed these experiments to show that the superior accuracy of our method compared to the other methods is due to the incorporation of the atlas information as we claim, rather than the differences in network architecture or training. Results show that the error of our method increases significantly when we discard the atlas information. When we only use the dMRI signal (last column in Table \ref{table:ablation}), the FA accuracy of our method is very close to that of Deep-DTI and Super-DTI (Table \ref{table:FA_results}) and the OD accuracy of our method is slightly worse than CNN-NODDI (Table \ref{table:OD_results}). Ablation experiments further show that all three extra pieces of information contribute to the model accuracy, but the contribution of atlas and atlas confidence is larger than the contribution of atlas-to-subject alignment error.

\begin{table*}[!htb]
\footnotesize
 \caption{\footnotesize{Results of some ablation experiments. In the column headings, we use $p$ to denote either FA or OD.}}
  \label{table:ablation}
   \begin{center}
    \begin{tabular}{ L{29mm}  C{25mm} C{21mm} C{21mm} C{19mm} }
\hline
input to the model & $\big[ s_k, \Phi_k(\mathbf{ \bar{p} }), $ $ \Phi_k(\mathbf{ p^{\sigma}}) , \Phi_k^{\text{err}} \big]$ & $\big[ s_k, \Phi_k(\mathbf{ \bar{p} }), $ $ \Phi_k(\mathbf{ p^{\sigma}}) \big]$  & $\big[ s_k, \Phi_k(\mathbf{ \bar{p} }) \big]$  & $s_k$    \\ \hline
FA, $n=6$  & $\bm{0.040 \pm 0.005}$ & $0.041 \pm 0.005$  & $0.044 \pm 0.006$ & $0.047 \pm 0.006$   \\ 
OD, $n=16$ & $\bm{0.047 \pm 0.004}$ & $0.050 \pm 0.004$  & $0.055 \pm 0.011$ & $0.060 \pm 0.010$   \\ 
\hline
\end{tabular}
  \end{center}
\end{table*}

\section{Conclusions}

We proposed a novel framework to incorporate atlases and DL for dMRI biomarker estimation. We cannot claim that the design of our framework is ``optimal''. For example, atlas-to-subject alignment may be improved by incorporating anatomical MRI information in addition to the diffusion tensor information. Nonetheless, our work has shown, for the first time, that spatio-temporal atlases can be used within a DL framework to achieve superior biomarker estimation accuracy from down-sampled data.

\section*{Acknowledgment}

This study was supported in part by the National Institute of Biomedical Imaging and Bioengineering and the National Institute of Neurological Disorders and Stroke of the National Institutes of Health (NIH) under award numbers R01EB031849, R01NS106030, and R01EB032366; in part by the Office of the Director of the NIH under award number S10OD0250111; in part by the National Science Foundation (NSF) under award 2123061; and in part by a Technological Innovations in Neuroscience Award from the McKnight Foundation. The content of this paper is solely the responsibility of the authors and does not necessarily represent the official views of the NIH, NSF, or the McKnight Foundation.

The DHCP dataset is provided by the developing Human Connectome Project, KCL-Imperial-Oxford Consortium funded by the European Research Council under the European Union Seventh Framework Programme (FP/2007-2013) / ERC Grant Agreement no. [319456]. We are grateful to the families who generously supported this trial.

\clearpage

\bibliographystyle{splncs04}
\bibliography{davoodreferences}

\end{document}